# Post-Quantum Cryptography-Based Bidirectional Authentication Key Exchange Protocol and Industry Applications: A Case Study of Instant Messaging

Abel C. H. Chen*, James W. H. Tung, Austin B. Y. Lin, Chin-Ling Chen, Ching-Chun Chang, Chin-Chen Chang

*Abstract*—This study aims to enhance the bidirectional authentication capability of ML-KEM (Module-Lattice-Based Key-Encapsulation Mechanism) by proposing the post-quantum cryptography-based (PQC-based) bidirectional authentication key exchange protocol. Furthermore, it introduces dual-usage certificates combining PQC-based DSA (Digital Signature Algorithm) and PQC-based KEM, which include composite schemes, catalyst schemes, and chameleon schemes. These dual-usage certificates utilize the PQC-based DSA public key and PQC-based KEM public key within the certificate to meet the requirements for bidirectional authentication and encryption, enabling the negotiation of a shared secret key. During the experimental phase, the study validates and compares key exchange message lengths and computation times under different certificate configurations. Finally, instant messaging is presented as an industry application to demonstrate the practical implementation of the proposed protocol.

## I. INTRODUCTION

IN August 2024, NIST finalized the standard documents for ML-KEM (Module-Lattice-Based Key-Encapsulation Mechanism)[1], ML-DSA (Module-Lattice-Based Digital Signature Algorithm) [2], and SLH-DSA (Stateless Hash-Based Digital Signature Algorithm)[3], making significant contributions to the transition to post-quantum cryptography. ML-KEM is particularly suitable for key encapsulation. For example, Bob can generate an AES-256 secret key and encrypt it using Alice's public key, producing a ciphertext to send to Alice. Alice can then decrypt the ciphertext with her private key to retrieve the AES-256 secret key. Subsequently, both parties can use the shared AES-256 secret key for encryption and decryption. This approach greatly enhances the construction of secure communications resistant to quantum computing attacks.

However, while ML-KEM provides robust key encapsulation mechanism, the described process does not authenticate the parties involved, as it lacks bidirectional authentication. To address this, this study proposes the post-quantum cryptography-based (PQC-based) bidirectional authentication key exchange protocol, which extends ML-KEM by incorporating bidirectional authentication. This ensures that only the communicating parties can derive the shared secret key. The main contributions of this study are as follows:

- Proposing a bidirectional authentication key exchange protocol that enables both parties to generate a shared secret key after mutual authentication.
- Proposing dual-usage certificates based on composite schemes [4], catalyst schemes [5] and chameleon schemes [6] combining PQC-based DSA (Digital Signature Algorithm) and PQC-based KEM (Key-Encapsulation Mechanism) to provide identity verification for the bidirectional authentication key exchange protocol.
- Conducting evaluations, including comparisons of message lengths and computational times, with a practical demonstration in the context of instant messaging applications.

## II. THE PROPOSED BIDIRECTIONAL AUTHENTICATION KEY EXCHANGE PROTOCOL AND APPLICATIONS

This section provides a detailed presentation of the proposed PQC-based bidirectional authentication key exchange protocol. The security aspects of the protocol are then analyzed. Finally, a case study on instant messaging is presented as an example of its application in an industry context.

### A. The Message Flows of the Proposed PQC-Based Bidirectional Authentication Key Exchange Protocol

The proposed PQC-based bidirectional authentication key exchange protocol, illustrated in Fig. 1, comprises a preliminary setup, three key exchange steps, and session key production, which are described as follows.

(1). **Preliminary Setup**
This study assumes that Alice possesses the PQC-based DSA key pair (including the private key $u_A$ and public key $U_A$), the PQC-based KEM key pair (including the private key

Abel C. H. Chen, James W. H. Tung, and Austin B. Y. Lin are with Information & Communications Security Laboratory, Chunghwa Telecom Laboratories, Taoyuan, Taiwan (e-mails: chchen.scholar@gmail.com, jamestung@cht.com.tw, and bylin@cht.com.tw)*(*Corresponding author: Abel C. H. Chen).*

Chin-Ling Chen is with the Department of Computer Science and Information Engineering, Chaoyang University of Technology, Taichung, Taiwan (e-mail: clc@cyut.edu.tw).

Ching-Chun Chang is with the National Institute of Informatics, Tokyo, Japan (e-mail: ccchang@nii.ac.jp).

Chin-Chen Chang is with the Department of Information Engineering, Feng Chia University, Taichung, Taiwan (e-mail: alan3c@gmail.com).

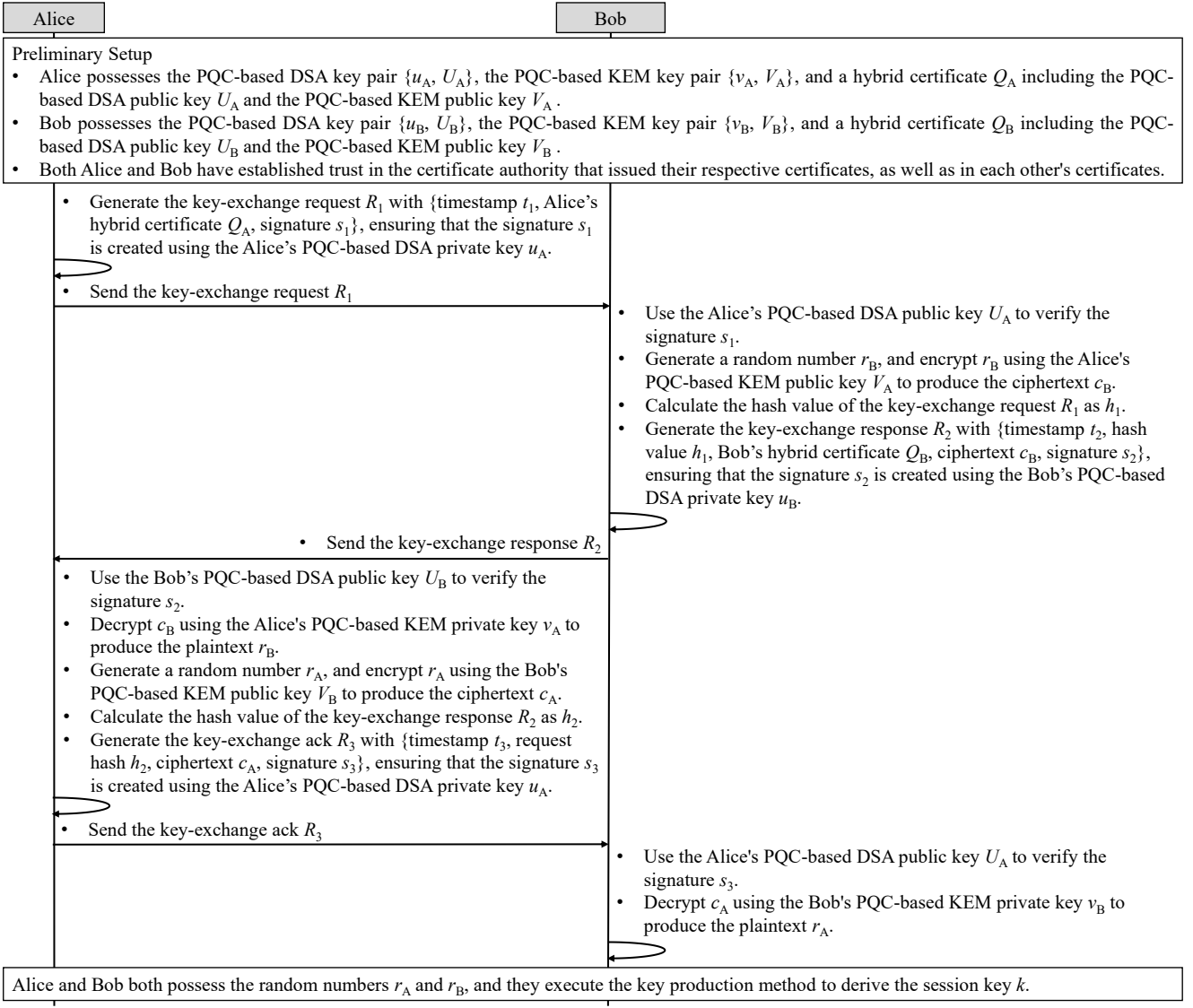

**Fig. 1.** The proposed post-quantum cryptography-based bidirectional authentication key exchange protocol.

$v_A$ and public key $V_A$), and a dual-usage certificate $Q_A$ (containing the PQC-based DSA public key $U_A$ and the PQC-based KEM public key $V_A$).

Similarly, Bob is assumed to have the PQC-based DSA key pair (including the private key $u_B$ and public key $U_B$), the PQC-based KEM key pair (including the private key $v_B$ and public key $V_B$), and a dual-usage certificate $Q_B$ (containing the PQC-based DSA public key $U_B$ and the PQC-based KEM public key $V_B$).

It is further assumed that both Alice and Bob trust the certificate authority (CA) that issued these dual-usage certificates and that all such certificates are valid.

### (2). Key Exchange Request

This study assumes that Alice acts as the initiator, beginning the process by sending the key exchange request $R_1$ to Bob. The key exchange request $R_1$ primarily includes the timestamp $t_1$, Alice's dual-usage certificate $Q_A$, and a signature $s_1$ generated by Alice using her PQC-based DSA private key $u_A$ to sign the request content. Subsequently, the key exchange request message $R_1$ is transmitted to Bob.

### (3). Key Exchange Response

Upon receiving the key exchange request $R_1$, Bob first extracts the dual-usage certificate $Q_A$ and verifies its validity. He then uses the PQC-based DSA public key $U_A$ contained within $Q_A$ to validate the signature $s_1$.

Once the verification is successful, Bob generates a random number $r_B$ and encrypts it using Alice's PQC-based KEM public key $V_A$ from $Q_A$, resulting in the ciphertext $c_B$. Additionally, Bob calculates the hash value of the key exchange request $R_1$ using a hash algorithm and extracts the last 8 bytes of the hash value, formatted as HashedId8, to derive $h_1$, which serves as the request ID.

Next, Bob constructs the key exchange response $R_2$, which includes the timestamp $t_2$, hash value $h_1$, Bob's dual-usage certificate $Q_B$, ciphertext $c_B$, and a signature $s_2$ generated by Bob using his PQC-based DSA private key $u_B$ to sign the response content. Finally, the key exchange response $R_2$ is transmitted to Alice.

### (4). Key Exchange Ack

Upon receiving the key exchange response $R_2$, Alice first extracts the certificate $Q_B$ and verifies its validity. She then uses the PQC-based DSA public key $U_B$ contained within $Q_B$ to validate the signature $s_2$.

Once the verification is successful, Alice generates a random number $r_A$ and encrypts it using Bob's PQC-based KEM public key $V_B$ from $Q_B$, producing the ciphertext $c_A$. Alice also calculates the hash value of the key exchange response $R_2$ using a hash algorithm and extracts the last 8 bytes of the hash value, formatted as HashedId8, to derive $h_2$, which serves as the response ID. Additionally, Alice utilizes her PQC-based KEM private key $v_A$ to decrypt the ciphertext $c_B$ and retrieve the random number $r_B$.

Subsequently, Alice constructs the key exchange ack $R_3$, which includes the timestamp $t_3$, hash value $h_2$, ciphertext $c_A$, and a signature $s_3$ generated by Alice using her PQC-based DSA private key $u_A$ to sign the acknowledgment content. Finally, the key exchange ack $R_3$ is transmitted to Bob.

Upon receiving the key exchange ack $R_3$, Bob uses the hash value $h_2$ to identify the corresponding key exchange response $R_2$ and retrieves Alice's certificate $Q_A$. He then utilizes the PQC-based DSA public key $U_A$ contained within $Q_A$ to validate the signature $s_3$. Once the verification is successful, Bob uses his PQC-based KEM private key $v_B$ to decrypt the ciphertext $c_A$ and retrieve the random number $r_A$.

### (5). Session Key Production

Following the aforementioned process, Alice and Bob have mutually authenticated each other's identities. Additionally, through the KEM mechanism, both parties securely share knowledge of the two random numbers $r_A$ and $r_B$. These random numbers can subsequently be used to generate the session key $k$. For example, an XOR (Exclusive-OR) operation could be performed directly on the two random numbers $r_A$ and $r_B$, which is the fastest method in terms of computation time. Alternatively, well-known and secure key derivation functions (KDFs) can be employed. However, since key derivation functions are not the primary focus of this study, they are not discussed in detail.

## B. Security Discussions

This subsection discusses the security of the proposed protocol from the following perspectives.

### (1). Identity Authentication

Both Alice and Bob possess certificates issued by a trusted CA. These certificates contain their respective PQC-based DSA public keys, which are considered trustworthy along with the certificates themselves. When Bob receives the key exchange request or the key exchange ack from Alice, he uses Alice's PQC-based DSA public key to verify the signatures within the messages. If the verification is successful, the messages can be trusted as originating from Alice. Similarly, when Alice receives the key exchange response from Bob, she uses Bob's PQC-based DSA public key to verify the signature. If the verification is successful, the message can be trusted as originating from Bob.

### (2). Data Confidentiality

The generated session key is based on the random numbers $r_A$ and $r_B$, making the confidentiality of these random numbers crucial.

The random number $r_A$ is generated by Alice and encrypted using Bob's PQC-based public key. If Bob possesses the corresponding PQC-based private key, he can decrypt the ciphertext to retrieve $r_A$. Since Bob's PQC-based public key is included in a certificate issued by the trusted CA, it can be considered reliable, ensuring that only Alice and Bob have access to $r_A$.

Similarly, the random number $r_B$ is generated by Bob and encrypted using Alice's PQC-based public key. If Alice possesses the corresponding PQC-based private key, she can decrypt the ciphertext to retrieve $r_B$. Since Alice's PQC-based public key is included in a certificate issued by the trusted CA, it can be considered reliable, ensuring that only Alice and Bob have access to $r_B$.

## C. Industry Applications: A Case Study of Instant Messaging

To provide a practical case study and application scenario, this study selects the instant messaging application developed by Chunghwa Telecom Laboratories as the case study. Before transmitting messages, the communicating parties can execute the proposed PQC-based bidirectional authentication key exchange protocol to establish a shared session key. This session key is then used to encrypt the transmitted messages, ensuring the security of their communication.

This section assumes that Alice (the initiator) and Bob intend to engage in instant messaging and first establish secure communication using the proposed PQC-based bidirectional authentication key exchange protocol. The following discussion focuses on the three key exchange steps—key exchange request, key exchange response, and key exchange acknowledgment—providing a detailed explanation of each.

### (1). Key Exchange Request

To construct the key exchange request $R_1$ as described in Section II.A.2, this study adopts the SignedData structure from the PKCS#7 format (shown in Fig. 2). Specifically, the ContentInfo field must be set with messageType as kepReq, and the message field should be set to NULL. It is important to note that this part remains subject to standardization. The constructed key exchange request $R_1$ will meet the aforementioned requirements and will contain the signature $s_1$, which is generated by Alice using her PQC-based DSA private key $u_A$ to sign the request content.

### (2). Key Exchange Response

To construct the key exchange response $R_2$ as described in Section II.A.3, this study adopts the SignedData structure from the PKCS#7 format (shown in Fig. 3). Specifically, the ContentInfo field must be set with messageType as kepResp, and the message field should contain the ciphertext $c_B$, which is generated by encrypting the random number $r_B$ using Alice's PQC-based KEM public key $V_A$. It is important to note that this part remains subject to standardization. The constructed key exchange response $R_2$ will meet the aforementioned requirements and will include the signature $s_2$, which is generated by Bob using his PQC-based DSA private key $u_B$ to sign the response content. Additionally, the response will contain the ciphertext $c_B$, which is derived by encrypting the random number $r_B$ using Alice's PQC-based KEM public key $V_A$.

```
SignedData (PKCS#7) (Key Exchange Request)
DigestAlgorithmIdentifiers: SHAKE-256
ContentInfo: {                          Need to be standardized
    messageType: kepReq
    message: NULL}
ExtendedCertificatesAndCertificates: {
    CA's certificate
    Alice's certificate}
SignerInfos: {
    SignerInfo: {
        Serial: ...
        DigestAlgorithmIdentifier: SHAKE-256
        SignedAttributes: {
            signingTime: ...
            messageDigest: ...}
        SignatureAlgorithmIdentifier: ML-DSA-65 (Alice)
        SignatureValue: ML-DSA-65 (Alice)    }
}
```

**Fig. 2.** Key exchange request in the PKCS#7 format.

```
SignedData (PKCS#7) (Key Exchange Response)
DigestAlgorithmIdentifiers: SHAKE-256
ContentInfo: {                          Need to be standardized
    messageType: kepResp
    message: ML-KEM Encapsulation
             (using Alice's public key)}
ExtendedCertificatesAndCertificates: {
    CA's certificate
    Bob's certificate}
SignerInfos: {
    SignerInfo: {
        Serial: ...
        DigestAlgorithmIdentifier: SHAKE-256
        SignedAttributes: {
            signingTime: ...
            messageDigest: ...}
        SignatureAlgorithmIdentifier: ML-DSA-65 (Bob)
        SignatureValue: ML-DSA-65 (Bob)    }
}
```

**Fig. 3.** Key exchange response in the PKCS#7 format.

(3). **Key Exchange Ack**

To construct the key exchange ack $R_3$ as described in Section II.A.4, this study adopts the SignedData structure from the PKCS#7 format (shown in Fig. 4). Specifically, the ContentInfo field must be set with messageType as kepAck, and the message field should contain the ciphertext $c_A$, which is generated by encrypting the random number $r_A$ using Bob's PQC-based KEM public key $V_B$. It is important to note that this part remains subject to standardization. The constructed key exchange ack $R_3$ will meet the aforementioned requirements and will include the signature $s_3$, which is generated by Alice using her PQC-based DSA private key $u_A$ to sign the acknowledgment content. Additionally, the acknowledgment will contain the ciphertext $c_A$, which is derived by encrypting the random number $r_A$ using Bob's PQC-based KEM public key $V_B$.

```
SignedData (PKCS#7) (Key Exchange Ack)
DigestAlgorithmIdentifiers: SHAKE-256
ContentInfo: {                          Need to be standardized
    messageType: kepAck
    message: ML-KEM Encapsulation
             (using Bob's public key)}
ExtendedCertificatesAndCertificates: {
    CA's certificate
    Alice's certificate}
SignerInfos: {
    SignerInfo: {
        Serial: ...
        DigestAlgorithmIdentifier: SHAKE-256
        SignedAttributes: {
            signingTime: ...
            messageDigest: ...}
        SignatureAlgorithmIdentifier: ML-DSA-65 (Alice)
        SignatureValue: ML-DSA-65 (Alice)    }
}
```

**Fig. 4.** Key exchange ack in the PKCS#7 format.

### III. EXPERIMENTAL RESULTS AND EVALUATION

To validate the proposed PQC-based bidirectional authentication key exchange protocol, this section adopts three dual-usage certificate schemes—Composite Scheme, Catalyst Scheme, and Chameleon Scheme—as well as a key exchange method that utilizes pure certificates for comparison. Accordingly, Section III.A provides an overview of the dual-usage certificate schemes, while Section III.B introduces the key exchange method based on pure certificates. Finally, Section III.C presents experimental data and discusses the results.

*A. Dual-usage Certificate Schemes*

This study focuses on integrating and comparing the following three dual-usage certificate schemes within the PQC-based bidirectional authentication key exchange protocol.

- Composite Scheme: The primary public key field in the certificate contains both the PQC-based DSA public key and the PQC-based KEM public key, with an additional OID (Object Identifier) included to describe this key combination.
- Catalyst Scheme: The primary public key field in the certificate contains the PQC-based DSA public key, while the PQC-based KEM public key is stored in the Alt. Public Key field within the Extensions section.
- Chameleon Scheme: The primary public key field in the certificate contains the PQC-based DSA public key, and the PQC-based KEM public key is stored in the Delta Certificate field within the Extensions section. It is important to note that the Delta Certificate is still a certificate and, therefore, must include the CA's signature to ensure its authenticity.

*B. The Compared Method Based on Pure Certificates*

The message flows of the compared method proposed in this study are illustrated in Fig. 5 and consists of three steps: key exchange request, key exchange response, and key exchange ack. The primary distinction between the compared method and the proposed PQC-based bidirectional authentication key exchange protocol lies in the type of certificates used. Specifically, the compared method utilizes pure certificates, whereas the PQC-based protocol employs dual-usage certificates.

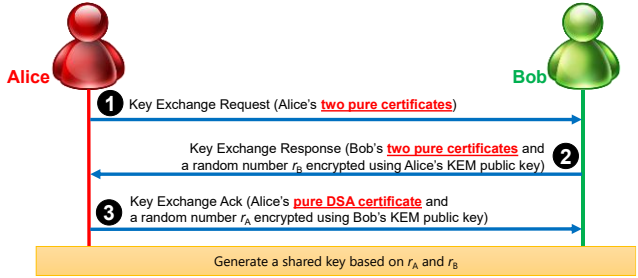

**Fig. 5.** The message flows of the compared method.

The key exchange process begins with Alice sending a key exchange request, which primarily includes her two pure certificates (a pure DSA certificate and a pure KEM certificate). Furthermore, Alice signs the key exchange request using her DSA private key.

Upon receiving and verifying the key exchange request, Bob encrypts a randomly generated value using the KEM public key from Alice's pure KEM certificate, producing a ciphertext. He then generates a key exchange response, signing it with his DSA private key. The key exchange response also includes Bob's two pure certificates (a pure DSA certificate and a pure KEM certificate).

Once Alice verifies the key exchange response, she encrypts a randomly generated value using the KEM public key from Bob's pure KEM certificate, producing a ciphertext. She then generates a key exchange ack and signs it with her DSA private key. It is worth noting that the key exchange ack only needs to include Alice's pure DSA certificate.

In the implementation process, each message in the proposed compared method can be structured using the PKCS#7 SignedData format, as illustrated in Fig. 2 to 4. The key distinction lies in the ExtendedCertificatesAndCertificates field, where multiple pure certificates are included.

*C. The Comparison of Message Lengths*

This section evaluates the proposed method using ML-DSA and SLH-DSA as the PQC-based DSA algorithms, along with ML-KEM as the PQC-based KEM algorithm. Three security levels are considered: Security Level 1, Security Level 3, and Security Level 5. The following discussion explores the different DSA approaches and parameter selections in detail.

(1). **The Proposed Protocol with ML-DSA**

The method combinations used for each security level in this section are as follows.

- Security Level 1: ML-DSA-44 + ML-KEM-512
- Security Level 3: ML-DSA-65 + ML-KEM-768
- Security Level 5: ML-DSA-87 + ML-KEM-1024

Table I presents the comparison results of the key exchange request. Based on the experimental findings, it can be observed that the proposed PQC-based bidirectional authentication key exchange protocol, which supports dual-usage certificates, requires only one certificate in the key exchange request. In contrast, the compared method requires two pure certificates, leading to a longer key exchange request length for the compared approach.

TABLE I
The Length Comparison of Key Exchange Request Based on ML-DSA (Unit: Bytes)

| Method | Security Level 1 | Security Level 3 | Security Level 5 |
|---|---|---|---|
| The Proposed Method with Composite Scheme | 7,549 | 10,351 | 14,011 |
| The Proposed Method with Catalyst Scheme | 7,575 | 10,377 | 14,037 |
| The Proposed Method with Chameleon Scheme | 10,050 | 13,741 | 18,719 |
| The Compared Method | 10,180 | 13,871 | 18,849 |

Furthermore, Composite Scheme and Catalyst Scheme exhibit shorter certificate lengths, resulting in a more noticeable reduction in the key exchange request length. On the other hand, the Chameleon Scheme, which includes a signature value in the Delta Certificate, does not significantly reduce the overall certificate length compared to the total length of two pure certificates. Consequently, the reduction in key exchange request length for the Chameleon Scheme is relatively limited.

Tables II and III present the comparison results for the key exchange response and key exchange ack, respectively. Since these two message formats are quite similar, their lengths are also nearly identical. The slight 2-byte difference is attributed to the fact that Alice's certificate's CommonName is 2 bytes longer than Bob's certificate's CommonName. Moreover, as the key exchange response and key exchange ack include a KEM-encrypted ciphertext in the message field, their lengths are longer than that of the key exchange request.

From the experimental results, it can be observed that because the proposed PQC-based bidirectional authentication key exchange protocol supports dual-usage certificates, only one certificate is required in the key exchange response, whereas the compared approach requires two pure certificates. As a result, the key exchange response in the proposed method is shorter compared to the compared method. However, in the key exchange ack, the compared method requires only one puare DSA certificate, making its message length shorter.

That said, when considering the total length of all three messages in the key exchange process, the proposed PQC-

based bidirectional authentication key exchange protocol demonstrates an overall advantage.

TABLE II
The Length Comparison of Key Exchange Response Based on ML-DSA (Unit: Bytes)

| Method | Security Level 1 | Security Level 3 | Security Level 5 |
|---|---|---|---|
| The Proposed Method with Composite Scheme | 8,325 | 11,447 | 15,587 |
| The Proposed Method with Catalyst Scheme | 8,351 | 11,473 | 15,613 |
| The Proposed Method with Chameleon Scheme | 10,826 | 14,837 | 20,295 |
| The Compared Method | 10,954 | 14,965 | 20,423 |

TABLE III
The Length Comparison of Key Exchange Ack Based on ML-DSA (Unit: Bytes)

| Method | Security Level 1 | Security Level 3 | Security Level 5 |
|---|---|---|---|
| The Proposed Method with Composite Scheme | 8,327 | 11,449 | 15,589 |
| The Proposed Method with Catalyst Scheme | 8,353 | 11,475 | 15,615 |
| The Proposed Method with Chameleon Scheme | 10,828 | 14,839 | 20,297 |
| The Compared Method | 7,511 | 10,249 | 14,005 |

(2). **The Proposed Protocol with SLH-DSA (small)**

The security levels in this section are based on the following method combinations:
- Security Level 1: SLH-DSA-128s + ML-KEM-512
- Security Level 3: SLH-DSA-192s + ML-KEM-768
- Security Level 5: SLH-DSA-256s + ML-KEM-1024

Tables IV and V present the comparison results for the key exchange request and key exchange response, respectively. Since the proposed PQC-based bidirectional authentication key exchange protocol is designed to support dual-usage certificates, only one certificate is required in both the key exchange request and key exchange response. In contrast, the compared method requires two pure certificates, resulting in a longer message length.

Furthermore, since SLH-DSA signatures are relatively long, and the compared method requires two separate signatures (one for each single-purpose certificate), the message length in the compared method is significantly affected, leading to a larger difference in message size. However, in the Chameleon Scheme, as the Delta Certificate still needs to include a signature, its results are more comparable to those of the compared method.

Table VI presents the comparison results for the key exchange ack. Since the compared method only includes a pure DSA certificate in the key exchange ack, its message length is shorter. However, when considering the total length across all three key exchange messages, the proposed PQC-based bidirectional authentication key exchange protocol, which incorporates dual-usage certificates, demonstrates a clear advantage.

TABLE IV
The Length Comparison of Key Exchange Request Based on SLH-DSA (small)(Unit: Bytes)

| Method | Security Level 1 | Security Level 3 | Security Level 5 |
|---|---|---|---|
| The Proposed Method with Composite Scheme | 17,107 | 34,275 | 61,811 |
| The Proposed Method with Catalyst Scheme | 17,131 | 34,299 | 61,835 |
| The Proposed Method with Chameleon Scheme | 25,042 | 50,578 | 91,685 |
| The Compared Method | 25,171 | 50,707 | 91,815 |

TABLE V
The Length Comparison of Key Exchange Response Based on SLH-DSA (small)(Unit: Bytes)

| Method | Security Level 1 | Security Level 3 | Security Level 5 |
|---|---|---|---|
| The Proposed Method with Composite Scheme | 17,883 | 35,371 | 63,387 |
| The Proposed Method with Catalyst Scheme | 17,907 | 35,395 | 63,411 |
| The Proposed Method with Chameleon Scheme | 25,818 | 51,674 | 93,261 |
| The Compared Method | 25,945 | 51,801 | 93,389 |

TABLE VI
The Length Comparison of Key Exchange Ack Based on SLH-DSA (small)(Unit: Bytes)

| Method | Security Level 1 | Security Level 3 | Security Level 5 |
|---|---|---|---|
| The Proposed Method with Composite Scheme | 17,885 | 35,373 | 63,389 |
| The Proposed Method with Catalyst Scheme | 17,909 | 35,397 | 63,413 |
| The Proposed Method with Chameleon Scheme | 25,820 | 51,676 | 93,263 |
| The Compared Method | 17,066 | 34,170 | 61,803 |

(3). **The Proposed Protocol with SLH-DSA (fast)**

The security levels in this section are based on the following method combinations:
- Security Level 1: SLH-DSA-128f + ML-KEM-512
- Security Level 3: SLH-DSA-192f + ML-KEM-768
- Security Level 5: SLH-DSA-256f + ML-KEM-1024

The comparison results for the key exchange request, key exchange response, and key exchange ack based on SLH-DSA (fast) are summarized in Tables VII, VIII, and IX, respectively. The experimental results are similar to those of SLH-DSA (small); however, since SLH-DSA (fast) has a longer signature value, the advantage of the proposed PQC-based bidirectional authentication key exchange protocol becomes more evident.

TABLE VII
The Length Comparison of Key Exchange Request Based on SLH-DSA (fast)(Unit: Bytes)

| Method | Security Level 1 | Security Level 3 | Security Level 5 |
|---|---|---|---|
| The Proposed Method with Composite Scheme | 35,571 | 73,158 | 101,942 |
| The Proposed Method with Catalyst Scheme | 35,595 | 73,182 | 101,966 |
| The Proposed Method with Chameleon Scheme | 52,738 | 108,903 | 151,879 |
| The Compared Method | 52,867 | 109,031 | 152,008 |

TABLE VIII
The Length Comparison of Key Exchange Response Based on SLH-DSA (fast)(Unit: Bytes)

| Method | Security Level 1 | Security Level 3 | Security Level 5 |
|---|---|---|---|
| The Proposed Method with Composite Scheme | 36,347 | 74,254 | 103,518 |
| The Proposed Method with Catalyst Scheme | 36,371 | 74,278 | 103,542 |
| The Proposed Method with Chameleon Scheme | 53,514 | 109,999 | 153,455 |
| The Compared Method | 53,641 | 110,125 | 153,582 |

TABLE IX
The Length Comparison of Key Exchange Ack Based on SLH-DSA (fast)(Unit: Bytes)

| Method | Security Level 1 | Security Level 3 | Security Level 5 |
|---|---|---|---|
| The Proposed Method with Composite Scheme | 36,349 | 74,256 | 103,520 |
| The Proposed Method with Catalyst Scheme | 36,373 | 74,280 | 103,544 |
| The Proposed Method with Chameleon Scheme | 53,516 | 110,001 | 153,457 |
| The Compared Method | 35,530 | 73,053 | 101,934 |

As the proposed protocol is designed with dual-usage certificates, each message requires one fewer signature compared to the compared method. This reduction in signature count leads to shorter message lengths, and this effect is even more pronounced when using SLH-DSA (fast) due to its inherently longer signature size.

(4). **Discussions**

In summary, the proposed PQC-based bidirectional authentication key exchange protocol, when combined with dual-usage certificates, reduces the number of certificates required in both the key exchange request and key exchange response, resulting in a shorter overall message length.

Furthermore, the reduction of one certificate not only leads to shorter message sizes but also eliminates the need for an additional signature verification process. As a result, the proposed protocol enhances computational efficiency, making it a more efficient alternative.

### IV. CONCLUSIONS AND FUTURE WORK

This study proposes a PQC-based bidirectional authentication key exchange protocol, which establishes bidirectional authentication resilient to quantum computing attacks and enables the generation of a shared session key for subsequent message encryption and decryption. Furthermore, this study proposes different PQC-based DSA and PQC-based KEM dual-usage certificate schemes, including composite schemes, catalyst schemes, and chameleon schemes, and compares the lengths of messages within the proposed PQC-based bidirectional authentication key exchange protocol across these dual-usage certificates.

In future research, the proposed PQC-based bidirectional authentication key exchange protocol can be applied to services that require bidirectional authentication. Furthermore, different DSA and KEM algorithms can be integrated to adapt the protocol for various applications.

### ACKNOWLEDGMENTS

The authors express their gratitude to NIST for organizing the KEM Workshop and for its contributions to advancing the global standardization of post-quantum cryptography. Appreciation is also expressed to the conference committee for their efforts in organizing and reviewing submissions, providing the opportunity to participate in discussions and knowledge exchange.